\documentclass[aps,twocolumn,amsmath,letterpaper]{revtex4}
\usepackage{amssymb}
\usepackage{graphicx}
\usepackage{array}
\usepackage{hhline}
\usepackage{longtable}

\renewcommand{\thetable}{\Roman{table}} \thetable

\begin{document}

\title{Quenched-Vacancy Induced Spin-Glass Order}
\author{G\"ul G\"ulp\i nar$^{1,3}$ and A. Nihat Berker$^{2,3,4}$}
\affiliation{$^1$Department of Physics, Dokuz Eyl\"ul University,
Buca 35160, Izmir, Turkey,} \affiliation{$^2$Department of Physics,
Ko\c{c} University, Sar\i yer 34450, Istanbul, Turkey,}
\affiliation{$^3$Feza G\"ursey Research Institute, T\"UB\.ITAK -
Bosphorus University, \c{C}engelk\"oy 34680, Istanbul, Turkey,}
\affiliation{$^4$Department of Physics, Massachusetts Institute of
Technology, Cambridge, Massachusetts 02139, U.S.A.}

\begin{abstract} The ferromagnetic phase of an Ising model in $d=3$, with any amount of quenched
antiferromagnetic bond randomness, is shown to undergo a transition
to a spin-glass phase under sufficient quenched bond dilution. This
general result, demonstrated here with the numerically exact global
renormalization-group solution of a $d=3$ hierarchical lattice, is
expected to hold true generally, for the cubic lattice and for
quenched site dilution. Conversely, in the
ferromagnetic-spinglass-antiferromagnetic phase diagram, the
spin-glass phase expands under quenched dilution at the expense of
the ferromagnetic and antiferromagnetic phases.  In the
ferro-spinglass phase transition induced by quenched dilution
reentrance as a function of temperature is seen, as previously found
for the ferro-spinglass transition induced by increasing the
antiferromagnetic bond concentration.

\end{abstract}
\maketitle

\def\s{\rule{0in}{0.28in}}
\setlength{\LTcapwidth}{\columnwidth}

\begin{figure}[h]
\includegraphics*[scale=1]{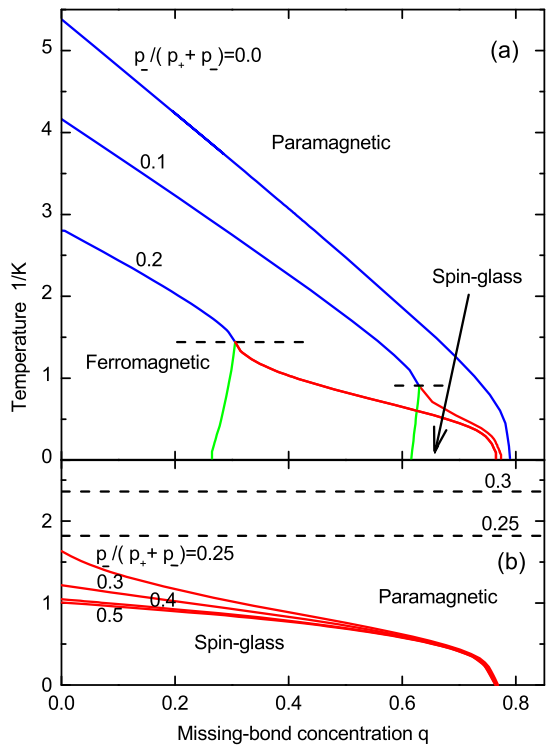}
\caption{(Color on-line) Calculated phase diagrams at constant
$p_-/(p_++p_-$). All phase transitions (full lines) are second
order. The dashed horizontal lines (shown only partly in (a)) are
not phase boundaries, but the Nishimori symmetry lines, given by
Eq.(\ref{eq:2}), for each value of $p_-/(p_++p_-$). The
multicritical points in (a), mediating the ferromagnetic,
spin-glass, and paramagnetic phases, occur on the Nishimori symmetry
lines.  These lines do not cross the phase boundaries at any other
type of point. Thus, the Nishimori symmetry lines in (b) are at
temperatures above the phase boundaries.}\label{fig1}
\end{figure}

The spin-glass phase \cite{NishimoriBook} is much studied due to its
prominent role in complex systems, as an example of random order. In
its simplest realization in the Ising model, the underlying system
has randomly distributed ferromagnetic and antiferromagnetic bonds.
In spatial dimension $d=3$, at low temperatures, ferromagnetic or
antiferromagnetic ordered phases occur when the system has
predominantly (\textit{e.g.}, more than $77\%$ \cite{Ozeki})
ferromagnetic or antiferromagnetic bonds, respectively. In between,
the spin-glass phase occurs.  The occurrence of the spin-glass
phase, in which the local degrees of freedom are frozen in random
directions, has strong implications in physical systems that are
realizations of the spin-glass system, spanning from materials
science to information theory and neural networks.

We have studied possibly the simplest modification of the spin-glass
system, to be commonly expected or realized in physical systems,
namely, the removal of bonds.  We find important qualitative and
quantitative effects. This Ising spin-glass system with quenched
bond vacancies \cite{Aharony, Giri, AharonyP, Southern, AharonyB,
Palmer, Bray, McKay} has the Hamiltonian
\begin{equation}\label{eq:1}
-\beta H = \sum_{\langle ij \rangle} K_{ij} s_i s_j\,,
\end{equation}
where $s_i = \pm 1$ at each site $i$ and $\langle i j \rangle$
indicates summation over nearest-neighbor pairs of sites.  The local
bond strengths are $K_{ij}=K>0$ with probability $p_+$, $K_{ij}=-K$
with probability $p_-$, or $K_{ij}=0$ with probability
$q=1-p_+-p_-$, respectively corresponding to a ferromagnetic
interaction, an antiferromagnetic interaction, or a bond vacancy.
This model has previously been studied at zero temperature
\cite{Southern, Bray} and in its spin-glass phase diagram
cross-section \cite{McKay} by position-space renormalization-group
theory, in $n$-replica version at zero-temperature \cite{Aharony,
AharonyP}, by mean-field theory \cite{Giri}, and by momentum-space
renormalization-group theory around $d=6$ dimensions \cite{Giri},
and by series expansion \cite{AharonyB, Palmer}.

We have performed the numerically exact renormalization-group
solution of this system on a $d=3$ hierarchical lattice
\cite{BerkerOstlund, Kaufman, Kaufman2}, to be given below.  Exact
solutions on hierarchical lattices constitute very good approximate
solutions for physical lattices.\cite{Erbas}  We calculate the
global phase diagram in the variables of temperature $1/K$, bond
vacancy concentration $q$, and antiferromagnetic bond fraction
$p_-/(p_+ + p_-)$, obtaining a rich structure and finding reentrance
as a function of temperature induced by bond vacancy.  Our results
agree with and extend the previous results \cite{Aharony, Giri,
AharonyP, Southern, AharonyB, Palmer, Bray, McKay} on this system.

Our results are most strikingly seen in Fig.\ref{fig1}.  The top
curve in Fig.1(a) corresponds to the quenched dilution of the system
with no antiferromagnetic bonds ($p_-=0$).  As the system is
quench-diluted, by increasing the missing-bond concentration $q$,
the transition temperature to the ferromagnetic phase is lowered
from its value with no missing bonds at $q=0$, until it reaches zero
temperature and the ferromagnetic phase disappears at the
percolation threshold of $q=0.789$ (to be compared with the value of
0.753 in the simple cubic lattice \cite{Essam}). However, with the
inclusion of even the smallest amount of antiferromagnetic bonds
(lower curves), a spin-glass phase, extending to finite
temperatures, always appears before percolation.  This result was
previously obtained at zero temperature \cite{Aharony, Southern,
AharonyB, Bray} and around $d=6$ \cite{Giri}.

The phase boundary between this vacancy-induced
ferromagnetic-spinglass phase transition shows reentrance, as also
seen \cite{Migliorini,Nobre,Roy} in conventional spin-glass phase
diagrams where the antiferromagnetic bond concentration is scanned.
In Fig.1(b), where the curves correspond to higher percentages of
antiferromagnetic interactions among the bonds present, starting
with $p_-/(p_++p_-) = 0.25$ in the top curve, the ferromagnetic
phase has disappeared and only spin-glass ordering occurs. As seen
from Fig.1(a), the percolation threshold of the spin-glass phase is
slightly lower than that of the pure ferromagnetic phase and, before
the disappearance of the ferromagnetic phase, the percolation
threshold of the spin-glass phase has a slight dependence on
$p_-/(p_++p_-)$.  The percolation threshold of the spin-glass phase
settles to the value of 0.763 after the disappearance of the
ferromagnetic phase.

Fig.2 shows the conventional phase diagrams of temperature versus
the fraction $p_-/(p_++p_-)$ of antiferromagnetic bonds in the
non-missing bonds, at fixed values of the dilution $q$.  As the
dilution is increased, the phases are depressed in temperature, as
can be expected. However, simultaneously, it is seen that the
spin-glass phase expands \cite{McKay} along the $p_-/(p_++p_-)$
axis, at the expense of the ferromagnetic and antiferromagnetic
phases, eventually dominating the entire low-temperature region.  It
is also seen, for $q=0.763$ and higher, that the spin-glass phase
occurs in the phase diagrams as two disconnected regions, near the
ferromagnetic and antiferromagnetic phases.  A similar disconnected
topology has recently been seen in the Blume-Emery-Griffiths spin
glass.\cite{Ozcelik}

\begin{figure}[h]
\includegraphics*[scale=1]{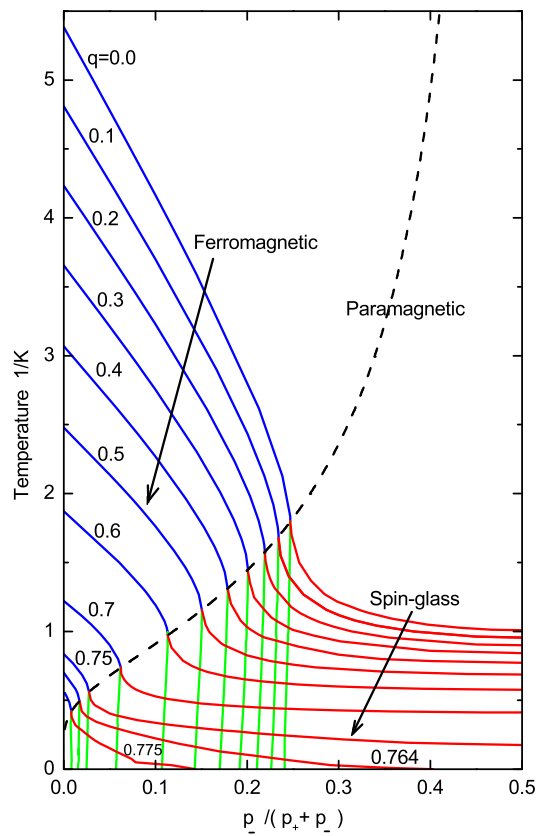}
\caption{(Color on-line) Calculated phase diagrams at constant
quenched dilution $q=1-p_++p_-$.  The phase diagrams being symmetric
about $p_-/(p_++p_-) = 0.5$, with the antiferromagnetic phase
replacing the ferromagnetic phase, only the $p_-/(p_++p_-) < 0.5$
halves are show.  All phase transitions (full lines) are second
order. The dashed curve is the Nishimori symmetry line given by
Eq.(\ref{eq:2}). All multicritical points, mediating the
ferromagnetic, spin-glass, and paramagnetic phases, lie on the
Nishimori symmetry line.}
\end{figure}

The dashed lines in Figs.1 and 2 are the Nishimori symmetry lines
\cite{NishimoriA, NishimoriB},
\begin{equation}\label{eq:2}
e^{\pm 2K} = \frac{p_+}{p_-}\,.
\end{equation}
All multicritical points occurring in the currently studied system
are on the Nishimori symmetry lines \cite{LeDoussal1, LeDoussal2},
as also previously seen \cite{McKay} for this system. Thus, as
illustrated in Fig.2, it is possible to continuously populate, with
multicritical points, the low-temperature segment of the Nishimori
line, by gradually changing the quenched dilution $q$. The Nishimori
symmetry condition appears in Fig.1 as a horizontal line for each
value of $p_-/(p_++p_-)$. This horizontal line intersects the upper
curve in Fig.1(a) at zero temperature, thereby implying the
occurrence of a zero-temperature multicritical point at the
percolation threshold, as also deduced from the sequence of phase
diagrams in Fig.2.  The horizontal lines of the Nishimori condition
intersect the two other phase diagrams in Fig.1(a) at their
multicritical point.  In Fig.1(b), multicritical points do not occur
and the horizontal lines of the Nishimori condition do not intersect
the phase boundaries, occurring at higher temperatures than the
phase boundaries.

\begin{figure}[h]
\includegraphics[scale=1.2]{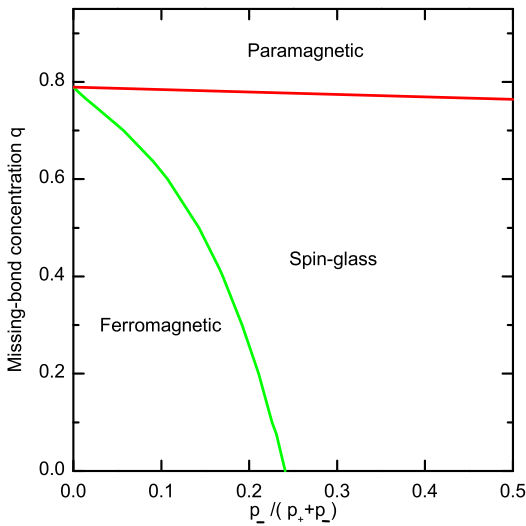}
\caption{(Color on-line) Calculated zero-temperature, $1/K = 0$,
phase diagram. All phase transitions are second order.}
\end{figure}

Fig.3 shows the zero temperature limit of the global phase diagram
of the currently studied system.  In the zero-temperature phase
diagram, it is again seen that a spin-glass phase intervenes
\cite{Aharony, Southern, AharonyB, Bray}, with smallest amount of
quenched antiferromagnetic bonds, between the ferromagnetic phase
and percolation, causing a direct ferromagnetic-spinglass phase
transition.

\begin{figure}.
\includegraphics[width=6cm]{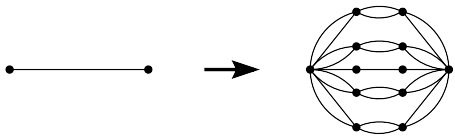}\\
\caption{The $d=3$ hierarchical lattice for which our calculation is
exact is constructed by the repeated imbedding of the graph as shown
in this figure. This hierarchical lattice gives numerically accurate
results for the critical temperatures of the $d =3$ isotropic and
anisotropic Ising models on the cubic lattice \cite{Erbas}.}
\end{figure}

Our method, detailed in other works \cite{Ozcelik, Falicov,
HinczBerker1, Guven}, will be briefly described now. We use the
$d=3$ hierarchical lattice whose construction is given in Fig.4.
This hierarchical lattice has the odd rescaling factor of $b=3$, for
the \textit{a priori} equivalent treatment of ferromagnetism and
antiferromagnetism, necessary for spin-glass problems.  Hierarchical
lattices admit exact solutions, by a renormalization-group
transformation that reverses the construction
steps.\cite{BerkerOstlund, Kaufman, Kaufman2}  Thus, hierarchical
lattices have become the testing grounds for a large variety of
cooperative phenomena, as also seen in recent works.\cite{Erbas,
Hinczewski2, Hinczewski3, ZZC, Khajeh, Rozenfeld, Kaufman3, Piolho,
Branco, Jorg, Boettcher, Monthus}. The hierarchical lattice of Fig.4
has been used in this work, because it gives numerically accurate
results for the critical temperatures of the $d =3$ isotropic and
anisotropic Ising models on the cubic lattice. \cite{Erbas}

In systems with quenched randomness, the renormalization-group
transformation determines the mapping of the quenched probability
distribution ${\cal P}(K)$.\cite{Andelman}  At each step, the
innermost unit of the lattice as pictured on the right side of Fig.4
is replaced by a single bond. This is effected by a series of
pairwise convolutions of the quenched probability distributions,
\begin{equation}\label{eq:3}
{\widetilde{{\cal P}}}(\widetilde{K}) = \int dK^I dK^{II} {\cal
P}_I(K^I){\cal P}_{II}(K^{II}) \delta(\widetilde{K} -
R(K^I,K^{II}))\,,
\end{equation}
\noindent where $R(K^I,K^{II})$ is
\begin{equation}\label{eq:4}
R(K_{ij}^I,K_{ij}^{II}) =  K_{ij}^I + K_{ij}^{II}
\end{equation}

\noindent for replacing two in-parallel random bonds with
distributions ${\cal P}_I(K^I)$ and ${\cal P}_{II}(K^{II})$ by a
single bond with ${\widetilde{{\cal P}}}(\widetilde{K})$, or
\begin{equation}\label{eq:5}
R(K_{ij}^I,K_{jk}^{II}) = \frac{1}{2} \ln \left[
\frac{\cosh(K_{ij}^I+K_{jk}^{II})}{\cosh(K_{ij}^I-K_{jk}^{II})}
\right]
\end{equation}

\noindent for replacing two in-series random bonds by a single bond.
The probability distributions are in the form of probabilities
assigned to interaction values, namely histograms.  Starting with
the three histograms described after Eq.(\ref{eq:1}), the number of
histograms quickly increases under the convolutions described above.
At a computational limit, a binning procedure is used before each
convolution to combine nearby histograms
\cite{Ozcelik,Falicov,HinczBerker1,Guven}, so that 160,000
histograms are kept to describe the probability distributions.  The
flows of these probability distributions, under iterated
renormalization-group transformations, determine the global phase
diagram of the system.

\begin{acknowledgments}
This research was supported by the Scientific and Technological
Research Council of Turkey (T\"UB\.ITAK), including computational
support through the TR-Grid e-Infrastructure Project hosted by
ULAKB\.IM, and by the Academy of Sciences of Turkey. G. G\"ulp\i nar
gratefully acknowledges the receipt of a B\.IDEP Postdoctoral
Fellowship from T\"UB\.ITAK.
\end{acknowledgments}

\end{document}